\documentclass[english,journal=jpccck,dvilaser,manuscript=article,layout=traditional]{achemso}
\usepackage[latin9]{inputenc}
\usepackage{textcomp}
\usepackage{amsmath}
\usepackage{amssymb}
\usepackage{stackrel}
\usepackage{wasysym}
\usepackage{esint}

\makeatletter

\title{Intermediate symmetric construction of transformation between anyon
and Gentile statistics }

\author{Yao Shen}

\affiliation{School of Criminal Investigation and Forensic Science, People's Public
Security University of China, Beijing 100038, PR China}

\email{shenyaophysics@hotmail.com}


\usepackage{amsfonts}
\usepackage{bbm}
\usepackage{epsfig}
\usepackage{achemso}
\usepackage{color}

\def\>{\rangle}
\def\<{\langle}

\makeatother

\usepackage{babel}
\begin{document}
\begin{abstract}
Gentile statistics describes fractional statistical systems in the
occupation number representation. Anyon statistics researches those
systems in the winding number representation. Both of them are intermediate
statistics between Bose-Einstein and Fermi-Dirac statistics. The second
quantization of Gentile statistics shows a lot of advantages. According
to the symmetry requirement of the wave function, we give the general
construction of transformation between anyon and Gentile statistics.
In other words, we introduce the second quantization form of anyons
in a easy way. Basic relations of second quantization operators, the
coherent state and Berry phase are also discussed.
\end{abstract}

\section{Introduction}

In nature, basic particles are divided into two categories: bosons
and fermions. Bosons obey Bose-Einstein statistics. Fermions follow
Fermi-Dirac statistics. The symmetry requests the wave functions of
bosons are symmetric, while they are anti-symmetric for fermions.
Regarding many body system, when we exchange two bosons, the wave
function is invariant. When two fermions are exchanged, the wave function
of the many body system gets a phase $\pi$ ($-1$).There could be
only one fermion in single quantum state at most, while the number
of boson in one state is not limited. In these decades, researchers
found some intermediate cases between Bose and Fermi statistics called
intermediate statistics or fractional statistics. Anyon and Gentile
statistics are two typical and important cases. Wilczek pointed out
that braiding two different anyons gave the wave function an additional
phase $2\pi k\alpha$, where $k$ is the winding number and $\alpha$
is statistical parameter \cite{x1,x2,x3,x4,x5,x6,x7,x8}. $\alpha$
could be fraction and depends on the type of anyon. For example, according
to the first Kitaev model ($1/2$ anyon ) \cite{x9,x10}, anyon has
four superselection sectors: $e$ (electric charge), $m$ (magnetic
vortex), $\varepsilon=e\times m$ and vacuum. Moving an anyon around
another which belongs to a different superselection sector except
vacuum gives the wave function the phase $\pi$ ($-1$). Gentile named
a special statistics after his name in 1940 \cite{x11,x12,x13,x14}.
The maximum occupation number of Gentile statistics is limited to
a finite number $n$\cite{p,q}. Because of these properties, the
representations of anyon and Gentile statistics are called the winding
number representation and the occupation number representation. They
describe intermediate statistics in different aspects. One of the
correspondences between these two representations has been discussed
in reference\cite{me}. That is a special algebra we constructed in
2010. More and more real anyon systems and Gentile systems are observed
these decades \cite{x16,x17,x18,x19,x20,x21,x24,x25}.

It has to be noticed that the wave function is also invariant when
one fermion goes around another. In another word, the wave function
of fermion goes back to itself for one circle. Exchanging two particles
is half a circle which gives fermion the phase factor $-1$ . Bosons
and fermions must be the ultimate limits of anyons and Gentile particles.
As for $1/2$ anyon, one circle corresponds to additional phase factor
$-1$, and half circle is a complex number $i$. In quantum mechanics,
all quantities can be observed are real. This limitation is inconvenient.
Fortunately, the group of anyon is braiding group. Therefore, it's
advisable that we can weaken the limitation of fermionic symmetry
to give a general construction of transformation between anyon and
Gentile statistics. In this case, the limits of anyons and Gentile
particles are bosons and fermions after adjustment.

This paper is organized as following: In section II, we construct
a general form of transformation between anyon and Gentile statistics.
In section III, we give the results of the second quantization operators
of anyons in the winding number representation. In section IV and
V, the coherent state and Berry phase of anyon are discussed respectively.
Finally, in section VI, the main results are concluded and further
work is discussed.

\section{The construction of transformation between two representations}

Gentile and anyon statistics are both intermediate statistics. The
maximum occupation number of Gentile particles in one state is a finite
number $n$. The state of Gentile statistics is represented to be
$\left|\nu\right\rangle _{n}$, where $\nu$ is the practical particle
number in one state. Thus the representation of Gentile statistics
is called the occupation number representation. $n\rightarrow\infty$
and $n=1$ correspond to Bose and Fermi statistics. It is very convenient
to describe systems in the occupation number representation. After
the second quantization, the basic operator relation is $\left[b,a^{\dagger}\right]_{n}=ba^{\dagger}-e^{i2\pi/\left(n+1\right)}a^{\dagger}b=1$,
where $a$, $b$, $a^{\dagger}$, $b^{\dagger}$ are annihilation
and creation operators of Gentile system and $a=b^{*}$. This relation
tells us that there will be a phase factor $e^{i2\pi/\left(n+1\right)}$
when two Gentile particles are exchanged . With regard to anyon, two
different anyons braid for one circle gives the phase factor $e^{i2\pi k\alpha}$.
We want to relate these two phase factors to each other. At the same
time, bosons and fermions are their limits. This is quite natural
for bosons. As we mentioned above, there is a problem for fermions.
Braiding two fermions for one circle gives the wave function nothing
while exchanging them gives $-1$. If we follow this requirement strictly,
for example, exchanging two different $1/2$ anyons will create a
complex number$i$. However, the difficulty lies in complex number
couldn't be observed in quantum mechanics. Considering these, we have
to weaken the limitation of symmetry. Because of the braiding group
of anyons, we assume that the wave functions of anyons go back to
themselves after braiding them for $g$ circles. Our basic assumption
goes to 
\begin{equation}
\alpha=\frac{1}{ng},k=\frac{\nu g}{2}.\label{eq:1}
\end{equation}
 $\nu$ is the practical particle number in one state and $\nu\in N$.
So we have the change of the winding number $\delta k=g/2$. The relation
of two phase factors between two representations is
\begin{equation}
e^{i2\pi k\alpha}=\left(e^{\frac{i2\pi\nu}{n+1}}\right)^{\frac{n+1}{2n}},\label{eq:2}
\end{equation}
or
\begin{equation}
e^{\frac{i2\pi\nu}{n+1}}=\left(e^{i2\pi k\alpha}\right)^{\frac{2}{1+\alpha g}}.\label{eq:3}
\end{equation}
In Bose case, $n\rightarrow\infty,\Rightarrow\alpha=0,k=\left(\nu g\right)/2$.
Equations (\ref{eq:2}) and (\ref{eq:3}) are satisfied automatically.
As for Fermi case $n=1$, we have $\alpha=1/g$ and $k=0$ ($\nu=0$)
or $k=g/2$ ($\nu=1$). There is no question when $k=0$ ($\nu=0$)
. When $k=g/2$ ($\nu=1$), both sides of equation (\ref{eq:2}) equal
to $-1$. In other words, this construction of transformation is successful
in its ultimate limits: bosons and fermions. What calls for special
attention is that reference \cite{me} is another example of $g=2$
(but reference \cite{me} shows a little difference). In reference
\cite{me}, $\alpha=1/(n+1)$ and $k=\nu$ ($k\in N$). It is a different
construction from this paper. Here we have $\alpha=1/(2n)$ and $k=\nu$
($k\in Q_{+}$). Both of these two constructions are right, because
the construction of transformation is not unique, as long as they
satisfy the symmetry requirement and self-consistent. 

\section{Basic relations of the second quantization operators}

The second quantization form has the advantage in describing the creation
and annihilation of particles. In this part, we transform the basic
relations of the second quantization operators in the occupation number
representation to the winding number representation to give anyon
another description.

According to equations (\ref{eq:1}) and (\ref{eq:3}), the main operator
relation becomes
\begin{equation}
\left[b,a^{\dagger}\right]_{\alpha,g}=ba^{\dagger}-\left(e^{i\pi\alpha g}\right)^{\frac{2}{1+\alpha g}}a^{\dagger}b=1.\label{eq:4}
\end{equation}
The states in the occupation number representation are $\left|\nu\right\rangle _{n}$.
In the winding number representation, the states can be expressed
as 
\begin{equation}
\left|k,g\right\rangle _{\alpha}\equiv\left|\frac{2k}{g}\right\rangle _{\alpha,g}=\left|\nu\right\rangle _{n}.\label{eq:5}
\end{equation}
And the variations of the practical occupation number and the winding
number are $\delta\nu=1$, $\delta k=g/2$ which means $\nu=0,1,2\cdots n$
and $k=0,g/2,g\cdots(ng)/2$. To create and annihilate a particle,
we have
\begin{equation}
a^{\dagger}\left|k,g\right\rangle _{\alpha}=\sqrt{\left\langle k+\frac{g}{2},g\right\rangle _{\alpha}}\left|k+\frac{g}{2},g\right\rangle _{\alpha},\label{eq:6}
\end{equation}
\begin{equation}
b\left|k,g\right\rangle _{\alpha}=\sqrt{\left\langle k,g\right\rangle _{\alpha}}\left|k-\frac{g}{2},g\right\rangle _{\alpha},\label{eq:7}
\end{equation}
where 
\begin{equation}
\left\langle k,g\right\rangle _{\alpha}\equiv\left\langle \frac{2k}{g}\right\rangle _{\alpha,g}=\frac{1-\left(e^{i2\pi k\alpha}\right)^{\frac{2}{1+\alpha g}}}{1-\left(e^{i\pi\alpha g}\right)^{\frac{2}{1+\alpha g}}}.\label{eq:8}
\end{equation}
$a^{\dagger}$ creates a particle each time and $b$ annihilates one.
And 
\begin{equation}
\left|k,g\right\rangle _{\alpha}=\frac{\left(a^{\dagger}\right)^{\frac{2k}{g}}}{\sqrt{\stackrel[p=g/2]{k}{\prod}\left\langle p,g\right\rangle _{\alpha}}}\left|0,g\right\rangle _{\alpha}=\frac{^{\left(b^{\dagger}\right)^{\frac{2k}{g}}}}{\sqrt{\stackrel[p=g/2]{k}{\prod}\left\langle p,g\right\rangle _{\alpha}^{*}}}\left|0,g\right\rangle _{\alpha}.
\end{equation}

We also construct the particle number operator $N$ as
\begin{equation}
N=\frac{1+\alpha g}{2\pi\alpha g}arccos\left[\frac{1}{2}\left(B+B^{\dagger}\right)\right],\label{eq:9}
\end{equation}
where
\begin{equation}
B\equiv ba^{\dagger}-a^{\dagger}b,\label{eq:10}
\end{equation}

Now, Let us consider the more general operator relations for arbitrary
operators: $u,v,w\cdots$. The intermediate statistical bracket -$n$
bracket is substituted by $\left[u,v\right]_{\alpha,g}$(see equation
(\ref{eq:4})). We give the famous Jacobi-like identities as examples,
\begin{equation}
\begin{array}{c}
\left[\left[u,v\right]_{\alpha,g},w\right]_{\alpha,g}+\left[\left[w,u\right]_{\alpha,g},v\right]_{\alpha,g}+\left[\left[v,w\right]_{\alpha,g},u\right]_{\alpha,g}\\
+\left[\left[v,u\right]_{\alpha,g},w\right]_{\alpha,g}+\left[\left[u,w\right]_{\alpha,g},v\right]_{\alpha,g}+\left[\left[w,v\right]_{\alpha,g},u\right]_{\alpha,g}\\
=\left(1-e\frac{i2\pi\alpha g}{1+\alpha g}\right)^{2}\left(uvw+wuv+vwu+vuw+uwv+wvu\right),
\end{array}\label{eq:11}
\end{equation}
\begin{equation}
\begin{array}{c}
\left[\left[u,v\right]_{\alpha,g},w\right]_{\alpha,g}+\left[\left[w,u\right]_{\alpha,g},v\right]_{\alpha,g}+\left[\left[v,w\right]_{\alpha,g},u\right]_{\alpha,g}\\
-\left[\left[v,u\right]_{\alpha,g},w\right]_{\alpha,g}-\left[\left[u,w\right]_{\alpha,g},v\right]_{\alpha,g}-\left[\left[w,v\right]_{\alpha,g},u\right]_{\alpha,g}\\
=\left(1-e\frac{i4\pi\alpha g}{1+\alpha g}\right)\left(uvw+wuv+vwu-vuw-uwv-wvu\right).
\end{array}\label{eq:12}
\end{equation}

\section{The coherent states }

The eigenstate of annihilation operator is called the coherent state
\cite{x26}. The coherent state of harmonic oscillator is very important
in physics. In Gentile statistics, the coherent state is the eigenstate
of annihilation operator $b$. In our construction, the coherent state
is expressed as $\left|\chi\right\rangle _{\alpha,g}$ in the winding
number representation, so we have another expression:
\begin{equation}
b\left|\chi\right\rangle _{\alpha,g}=\chi\left|\chi\right\rangle _{\alpha,g},\label{eq:13}
\end{equation}
where the eigenvalue $\chi$ is a Grassmann number. Grassmann numbers
don't commute with the states. They satisfy
\begin{equation}
\chi\left|k,g\right\rangle _{\alpha}=\lambda_{\alpha}(k,g)\left|k,g\right\rangle _{\alpha}\chi,\label{eq:14}
\end{equation}
here we take
\begin{equation}
\lambda_{\alpha}(k,g)\equiv\left(e^{\pm i2\pi k\alpha}\right)^{\frac{2}{1+\alpha g}},\label{eq:15}
\end{equation}
as an example and we also have
\begin{equation}
\chi^{\frac{1+\alpha g}{\alpha g}}=0.\label{eq:16}
\end{equation}
Under these assumptions, we can construct the coherent state as
\begin{equation}
\left|\chi\right\rangle _{\alpha,g}=M\stackrel[k=0]{\frac{1}{2\alpha}}{\sum}\gamma_{\alpha}\left(k,g\right)\left|k,g\right\rangle _{\alpha}\chi^{\frac{2k}{g}},\label{eq:17}
\end{equation}
where $M$is the normalization constant, the coefficients read
\begin{equation}
\gamma_{\alpha}\left(k,g\right)=\stackrel[p=0]{k-g/2}{\prod}\frac{e^{\pm\frac{i4\pi p\alpha}{1+\alpha g}}}{\sqrt{\left\langle p,g\right\rangle _{\alpha}}},\label{eq:18}
\end{equation}
and both the variations $\delta k=\delta p=g/2$. In this case, the
state in adjoint space is 
\begin{equation}
\left\langle \chi\right|_{\alpha,g}=M\stackrel[k=0]{\frac{1}{2\alpha}}{\sum}\gamma_{\alpha}^{*}\left(k,g\right)\bar{\chi}^{\frac{2k}{g}}\left\langle k,g\right|_{\alpha},\label{eq:19}
\end{equation}
where $\bar{\chi}$ is also the Grassmann number in adjoint space.
Under the normalization condition $\left\langle \chi|\chi\right\rangle _{\alpha,g}=1$,
we have
\begin{equation}
M=\left[1+\stackrel[l=g/2]{\frac{1}{2\alpha}}{\sum}\left(\bar{\chi}\chi\right)^{\frac{2l}{g}}\left|\gamma_{\alpha}\left(l,g\right)\right|^{2}\right]^{-\frac{1}{2}},\label{eq:20}
\end{equation}
and also the variation $\delta l=g/2$. Moreover, the relations of
the Grassmann number and creation and annihilation operators can be
obtained :
\begin{equation}
\chi\left(a^{\dagger}\right)^{\frac{2k}{g}}=\left(e^{\pm i2\pi k\alpha}\right)^{\frac{2}{1+\alpha g}}\left(a^{\dagger}\right)^{\frac{2k}{g}}\chi,
\end{equation}
\begin{equation}
a^{\frac{2k}{g}}\chi=\left(e^{\pm i2\pi k\alpha}\right)^{\frac{2}{1+\alpha g}}\chi a^{\frac{2k}{g}}.
\end{equation}
As for $b^{\dagger},b$, there will be the same relations.

\section{Berry phase}

Berry phase is a very important concept in differential geometry.
The appearance of Berry phase is a topological effect. The source
is that the spherical surface is not homeomorphic to plane. Under
an adiabatic and unitary transformation, the eigenstate of Gentile
system could obtain a phase factor $exp(i\eta_{\nu})$ where $\eta_{\nu}$
is an angle which depends on the practical particle number, and this
factor is Berry phase. We assume the unitary transformation is denoted
by $U\left(\theta\right)$, where $\theta$ is some kind of angle.
Operator $A$ under this unitary transformation become $U\left(\theta\right)AU^{\dagger}\left(\theta\right)$.
In the occupation number representation of Gentile statistics, we
adopt
\begin{equation}
U\left(\theta\right)=e^{-i\theta J_{z}},
\end{equation}
and $J_{z}=N=n/2$. We have 
\begin{equation}
\eta_{\nu}=i\varoint\left\langle \nu|U^{\dagger}(\theta)\nabla_{\theta}U(\theta)\left|\nu\right\rangle _{n}\right.=2\pi(\nu-\frac{n}{2}).\label{eq:24}
\end{equation}
According to equation (\ref{eq:1}), we relate $exp(i\eta_{\nu})$
to the phase factor of anyon in the winding number representation,
equation (\ref{eq:24}) becomes 
\begin{equation}
\eta_{\nu}=2\pi(\frac{2k}{g}-\frac{1}{2\alpha g}).\label{eq:26}
\end{equation}
The positive and negative value of $\eta_{\nu}$ correspond to two
closed trajectories in two opposite directions.

To show the non-uniqueness of the construction, here we introduce
another instruction, suppose $\alpha=1/(ng)$ and 
\begin{equation}
e^{i\eta_{\nu}}=e^{i2\pi k\alpha}.\label{eq:27}
\end{equation}
Then we get a restrict of the winding number
\begin{equation}
k=\frac{\nu}{\alpha}-\frac{1}{2\alpha g}.\label{eq:28}
\end{equation}
The positive and negative value of $k$ means braiding clockwise and
counterclockwise. For instance, when $n=3$ and $g=2$, we have $\nu=0,1,2,3$
and $\alpha=1/6$. So the restrict of $k$ is $k=6\nu-9=-9,-3,3,9$.
The difference is that equation (\ref{eq:26}) is a direct usage of
equation (\ref{eq:1}), and equation (\ref{eq:27}) forces those two
phase factors equal to each other to give a restrict of the winding
number.

\section{Conclusion and discussion}

Anyon and Gentile statistics are two typical intermediate statistics
beyond Bose-Einstein and Fermi-Dirac statistics. Bose and Fermi statistics
are their ultimate limit conditions. When two different kinds of anyons
are braiding, the wave function of the system could obtain an additional
phase factor which depends on the statistical parameter and the winding
number. The maximum occupation number of Gentile statistics is neither
$1$ nor $\infty$, but a finite number $n$. After the second quantization,
exchanging two Gentile particles also gives the wave function a phase
factor which depends on $n$. The second quantization form is very
convenient to create, annihilate particles and gives special properties
of the system. So it is worth researching the second quantization
form of anyons. But this is not easy and the symmetry requirement
must be weakened to make the construction self-consistent. In this
paper, we introduce a general construction of transformation between
anyon and Gentile statistics to change the results of Gentile statistics
in the winding number representation of anyons. In other words, we
give the second quantization form of anyons indirectly. The coherent
state and Berry phase are also discussed. Further researches of anyon
properties are proceeding.

\section{Acknowledgment}

The research was supported by the Fundamental Research Funds for the
Central Universities No.2020JKF306.


\begin{thebibliography}{References}
\bibitem{x1} F. Wilczek, Phys. Rev. Lett. \textbf{49,} 957 (1982).

\bibitem{x2} F. Wilczek, Phys. Rev. Lett. \textbf{48,} 1144 (1982).

\bibitem{x3} F. Wilczek, Phys. Rev. Lett. \textbf{48,} 1146 (1982).

\bibitem{x4} S. Levit, and N. Sivan, Phys. Rev. Lett. \textbf{69,}
363 (1992).

\bibitem{x5} A. Lerda (Ed.), Anyons-Quantum Mechanics of Particles
with Fractional Statistics (Springer-Verlag, Berlin Heidelberg, 1992)
and references therein.

\bibitem{x6} A. Comtet, S.N. Majumdar, and S. Ouvry, J. Phys. A \textbf{40,}
11255 (2007).

\bibitem{x7} S. Mashkevich, S. Matveenko, and S. Ouvry, Nuclear Phys.
B \textbf{763,} 431 (2007).

\bibitem{x8} A. Rovenchak, Low Temp. Phys. \textbf{35,} 400 (2009).

\bibitem{x9} A.Y. Kitaev, Ann. Phys. \textbf{303,} 2 (2003).

\bibitem{x10}A. Kitaev, Ann. Phys. \textbf{321,} 2 (2006).

\bibitem{x11}G. Gentile, Nuovo Cimento \textbf{17, }493 (1940).

\bibitem{x12} A. Khare, Fractional Statistics and Quantum Theory
(World Scientific, Singapore, 1997).

\bibitem{x13}R. Hernandez-Perez, and D. Tun, Physica A \textbf{384,}
297304 (2007).

\bibitem{x14}W.S. Dai, and M. Xie, J. Stat. Mech. P07034 (2009).

\bibitem{p}Y. Shen, W.S. Dai, and M. Xie, Phys. Rev. A \textbf{75,}
042111 (2007).

\bibitem{q}W.S. Dai, and M. Xie, Physica A\textbf{ 331,} 497(2004).

\bibitem{me} Y. Shen, Q. Ai, and G. L. Long. Physica A. \textbf{389,}
1565-1570 (2010).

\bibitem{x16}Y. Shen, and Y. L. Zhao. J. Phys. Chem. A. \textbf{122},
6349\textminus 6353 (2018).

\bibitem{x17}W.S. Dai, and M. Xie, J. Stat. Mech. P04021 (2009).

\bibitem{x18}B. DeMarco, and D.S. Jin, Science \textbf{285,} 1703
(1999).

\bibitem{x19}A.G. Truscott, K.E. Strecker, W.I. McAlexander, G.B.
Partridge, and R.G. Hulet, Science \textbf{291,} 2570 (2001).

\bibitem{x20}F. Schreck, L. Khaykovich, K.L. Corwin, G. Ferrari,
T. Bourdel, J. Cubizolles, and C. Salomon, Phys. Rev. Lett. \textbf{87,}
080403 (2001).

\bibitem{x21}F. Wilczek (Ed.), Fractional Statistics and Anyon Superconductivity
(World Scientific, Singapore, 1990) and references therein.

\bibitem{x24}Z.N.C. Ha, Phys. Rev. Lett. \textbf{73, }1574 (1994).

\bibitem{x25}B. Sutherland, Phys. Rev. B \textbf{56,} 4422(1997).

\bibitem{x26}J.R. Klauder, and B.-S. Skagerstam, Coherent States,
Applications in Physics and Mathematical Physics (World Scientific,
Singapore, 1985).
\end{thebibliography}
\end{document}